\documentclass[8.5pt,twoside,twocolumn]{article}
\usepackage[super,sort&compress,comma]{natbib} 
\usepackage[version=3]{mhchem}
\usepackage{balance}
\usepackage{times,mathptm}
\usepackage{sectsty}
\usepackage{graphicx} 
\usepackage{lastpage}
\usepackage[format=plain,justification=raggedright,singlelinecheck=false,font=small,labelfont=bf,labelsep=space]{caption} 
\usepackage{fancyhdr}
\newcommand{\hext}{H_\mathrm{ext}}
\newcommand{\hres}{H_\mathrm{res}}
\newcommand{\fres}{f_\mathrm{res}}

\newcommand{\lmesh}{l_{\mathrm{mesh}}}

\begin{document}

\thispagestyle{plain}
\fancypagestyle{plain}{

\renewcommand{\headrulewidth}{1pt}}
\renewcommand{\thefootnote}{\fnsymbol{footnote}}
\renewcommand\footnoterule{\vspace*{1pt}%
\hrule width 3.4in height 0.4pt \vspace*{5pt}} 
\setcounter{secnumdepth}{5}

\makeatletter 
\renewcommand\@biblabel[1]{#1}            
\renewcommand\@makefntext[1]%
{\noindent\makebox[0pt][r]{\@thefnmark\,}#1}
\makeatother 
\renewcommand{\figurename}{\small{Fig.}~}
\sectionfont{\large}
\subsectionfont{\normalsize} 

\fancyfoot{}

\fancyhead{}
\renewcommand{\headrulewidth}{1pt} 
\renewcommand{\footrulewidth}{1pt}
\setlength{\arrayrulewidth}{1pt}
\setlength{\columnsep}{6.5mm}
\setlength\bibsep{1pt}

\twocolumn[
  \begin{@twocolumnfalse}
\noindent\LARGE{\textbf{Sensing magnetic nanoparticles using nano-confined ferromagnetic resonances in a magnonic crystal}}
\vspace{0.6cm}

\noindent\large{\textbf{Peter J.~Metaxas,$^{\ast}$\textit{$^{a\ddag}$} Manu Sushruth,\textit{$^{a\ddag}$} Ryan A.~Begley,\textit{$^{a}$} Junjia Ding,\textit{$^{b\textsection}$} Robert C.~Woodward,\textit{$^{a}$} Ivan S.~Maksymov,\textit{$^{a}$} Maximilian Albert,\textit{$^{c}$} Weiwei Wang,\textit{$^{c}$} Hans Fangohr,\textit{$^{c}$}  Adekunle O.~Adeyeye\textit{$^{b}$} and
Mikhail Kostylev\textit{$^{a}$}}}\vspace{0.5cm}

 \end{@twocolumnfalse} \vspace{0.6cm}

  ]

\noindent\textbf{We demonstrate the use of the magnetic-field-dependence of highly spatially confined, GHz-frequency ferromagnetic resonances in a ferromagnetic nanostructure for the detection of adsorbed magnetic nanoparticles. This is achieved in a large area magnonic crystal consisting of a thin ferromagnetic film containing a periodic array of closely spaced, nano-scale anti-dots. Stray fields from nanoparticles within the anti-dots modify  resonant dynamic magnetisation modes in the surrounding magnonic crystal, generating easily measurable resonance peak shifts. The shifts are comparable to the resonance linewidths for high anti-dot filling fractions with their signs and magnitudes dependent upon the modes' localisations (in agreement with micromagnetic simulation results). This is a highly encouraging result for the development of frequency-based nanoparticle detectors for high speed nano-scale biosensing.}
\section*{}
\vspace{-1cm}

\footnotetext{\textit{$^{a}$~School of Physics, M013, University of Western Australia, 35 Stirling Hwy, Crawley WA 6009, Australia. Fax: +61 8 6488 7364; Tel: +61 8 6488 7015; E-mail: peter.metaxas@uwa.edu.au}}
\footnotetext{\textit{$^{b}$~Information Storage Materials Laboratory, Department of Electrical and Computer Engineering, National University of Singapore, Singapore-117576, Singapore. }}
\footnotetext{\textit{$^{c}$~Engineering and the Environment, University of Southampton, Southampton, SO17 1BJ, United Kingdom. }}
\footnotetext{{$^{\textsection}$~Present address: \textit{Argonne National Laboratory, 9700 S. Cass Avenue, Argonne, IL 60439, USA.}}}
\footnotetext{{$^{\ddag}$~These authors contributed equally to this work.}}

Magnetic biosensors, in which biological analytes are tagged with magnetic nanoparticles (MNPs), have excellent potential for solid-state point-of-care medical diagnostics \cite{Gaster2009,Srinivasan2011,Freitas2012}. The technique is intrinsically matrix-insesntive\cite{Gaster2009}, can compete with industry-standard immunoassays\cite{Srinivasan2009} and  can  be combined with  magnetic separation methods\cite{Osterfeld2008}. The  central element of a magnetic biosensor is a detector for the stray or `fringing' magnetic fields generated by magnetised MNPs which are used to label, typically \textit{in-vitro},  analytes of interest within a biological sample. Previously used sensors include SQuIDs \cite{Chemla2000}, Hall sensors\cite{Besse2002}, ferromagnetic rings\cite{Miller2002,Llandro2007} and  magneto-impedance devices\cite{Devkota2013}. However one of the most widely used methods is that employing  magnetoresistive (MR) magnetic field sensors \cite{Gaster2009,Baselt1998,Osterfeld2008,Srinivasan2009,Hall2010,Martins2010,Srinivasan2011,Freitas2012,Li2014} which are typically fabricated with at least one lateral dimension on the order of 10-100 $\mu$m. An MNP is detected when its stray (or `fringing') magnetic field modifies the quasi-static magnetic configuration in the ferromagnetic MR device. This changes the device's resistance, enabling electronic MNP detection.  

It can however be challenging to minimise noise in conventional MR sensors when  reducing the sensor size due to thermal instabilities of the device's magnetic configuration\cite{Nor1998,Smith2006,Lei2011}. A suggested approach to overcome this is to use intrinsically high frequency detection methods exploiting the strong field dependence of resonant  magnetisation dynamics\cite{Braganca2010,Mizushima2010,Inoue2011} which can be reliably driven in isolated nanostructures. These dynamics can be driven electrically in spin torque oscillators (e.g.~\cite{Kiselev2003,Rippard2004,Braganca2010,Kim2012a,Locatelli2013} and refs.~therein) which have been predicted to retain high field sensing signal to noise ratios at sub-100 nm dimensions\cite{Braganca2010,Mizushima2010}. Furthermore, real time electrical detection of the  dynamics\cite{Krivorotov2005,Suto2010,Zeng2012a} will pave the way for  high speed\cite{Mizushima2010}, nano-scale MNP sensing for solid-state flow cytometry\cite{Loureiro2009,Loureiro2011,Loureiro2011a,Helou2013}.

\begin{figure}[htbp]
\centering
	\includegraphics[width=7cm]{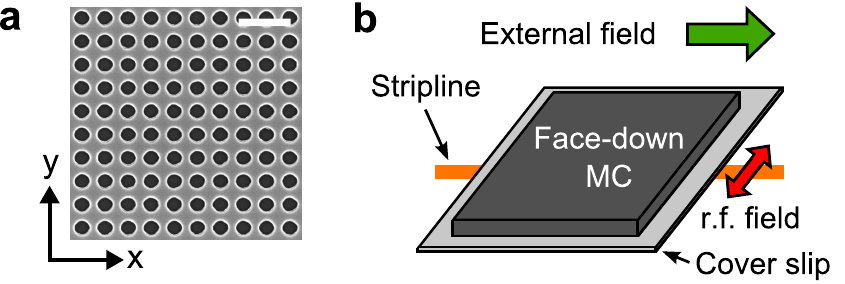}
	\caption{(a) Scanning electron micrograph of a portion of a magnonic crystal (MC). The scale bar is 1 $\mu$m long. (b) Schematic of the experimental setup showing the MC placed face down above a micro-stripline. }
	\label{fig1}
\end{figure}

In this work we use a large area\cite{Singh2004,Adeyeye2008}  magnonic crystal (MC)\cite{Krawczyk2014}  to macroscopically probe resonant GHz-frequency magnetisation dynamics which are spatially confined to nanometric regions and experimentally demonstrate their use for MNP sensing.   In contrast to continuous ferromagnetic layers, nanostructured MCs (e.g.~Fig.~\ref{fig1}a) exhibit a number of distinct ferromagnetic resonance (FMR) modes with different lateral spatial localisations within the crystal's nano-periodic structure\cite{Neusser2008,Tacchi2010,Zivieri2013}.  A previous study demonstrated that FMR modes within anti-dot-based MCs are sensitive to magnetic nanostructures fabricated within the anti-dots, leading to a proposal for a magnonic biosensor to detect captured MNP-based tags\cite{Ding2013}.  Here we demonstrate that stray magnetic fields generated by captured MNPs within an anti-dot-based MC do indeed generate clear resonance peak shifts which are mode-dependent and which approach the resonance linewidth when the fraction of MNP-filled anti-dots is high. Note that resonances within MNPs can be be detected directly however only broad, relatively weak signals have been observed previously for small collections of MNPs\cite{Chatterjee2011}.  Rather, in this work, we detect MNP-induced changes to very well defined resonances within a periodically nanostructured, high quality ferromagnetic film.

Large area MCs ($4 \times 4$ mm$^2$) consisting of arrays of 0.3 $\mu$m wide anti-dots with edge-to-edge spacings of 0.15 $\mu$m were fabricated  in a 30 nm thick Ni$_{80}$Fe$_{20}$ ferromagnetic layer using deep ultraviolet lithography\cite{Singh2004,Adeyeye2008} (Fig.~\ref{fig1}a). The MCs' FMR modes were probed at room temperature using broadband, stripline-based, magnetic field modulated FMR spectroscopy (Fig.~\ref{fig1}b). This technique  measures the derivative of finite width FMR peaks with respect to the external magnetic field, $\hext$, at a fixed GHz frequency. MC spectra are measured both before and after the addition of cluster-shaped  MNPs (diameters $\sim 0.1-0.3$ $\mu$m). Two  micromagnetic simulation methods have also been used: (i) a  time domain (`ringdown') simulation using MuMax3\cite{Vansteenkiste2014} which subjects the MC to a field pulse and exploits Fourier analysis on the resulting dynamics to extract frequency-resolved information (e.g.~\cite{Grimsditch2004}), and; (ii) an  eigenmode method (e.g.~\cite{dAquino2009}) which directly calculates the system's resonant modes and mode profiles for a given $\hext$ (carried out using FinMag which is the successor to Nmag\cite{Fischbacher2007}).  See the supplementary information for additional details.

\begin{figure}[htbp]
\centering
	\includegraphics[width=8cm]{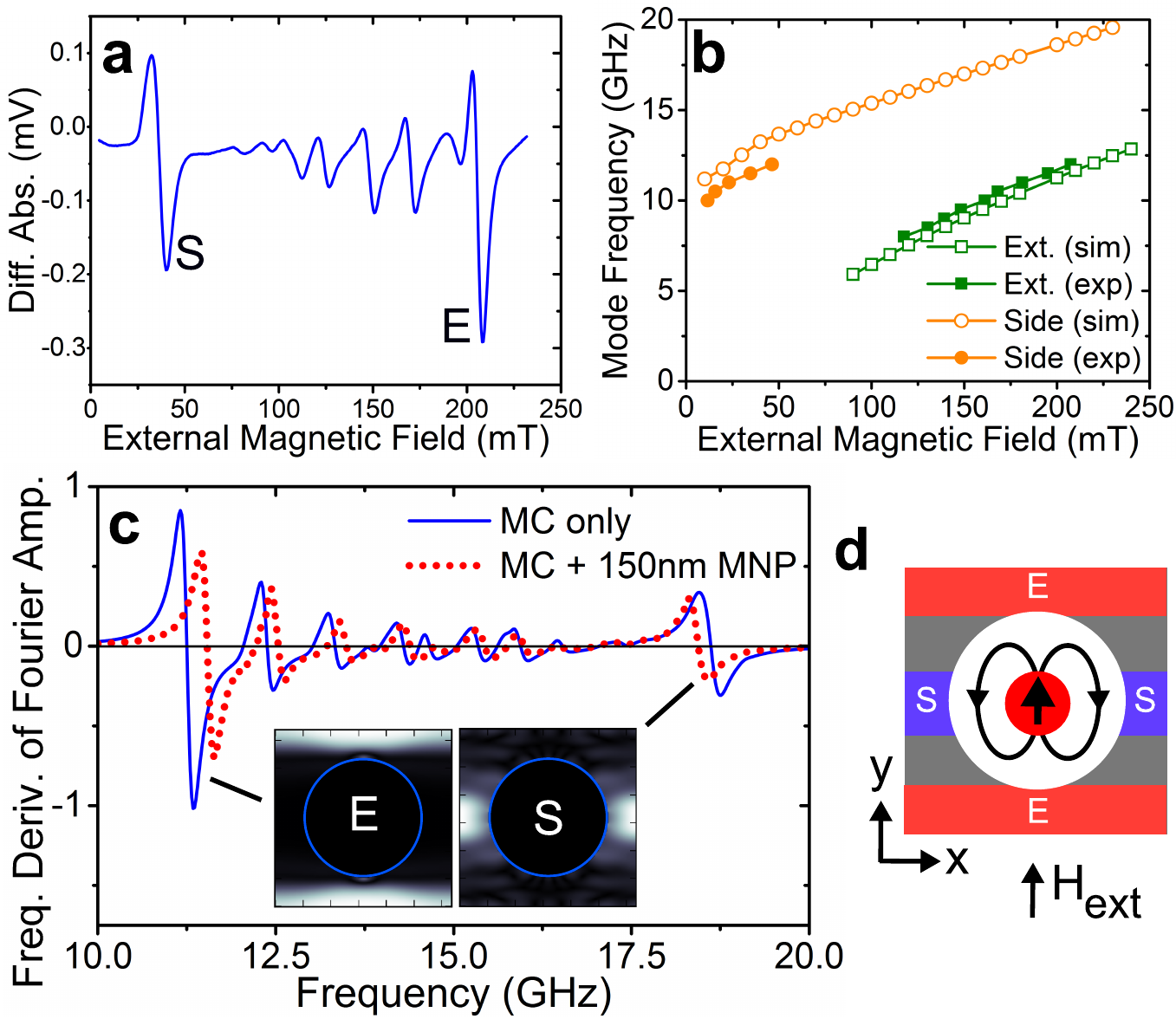}
	\caption{(a) Experimental, field-resolved FMR trace ($f=11.5$ GHz)  showing differential absorption peaks corresponding to  FMR modes in the MC. (b) Comparison of experimental and simulated resonant frequencies versus $\mu_0\hext$ for the side (S) and extended (E) modes. (c) Frequency-resolved, Fourier transformed time domain simulation data at $\mu_0 \hext=200$ mT for a single unit cell of the MC with and without a 150 nm MNP at the centre of the anti-dot.  The Fourier amplitude has been differentiated with respect to $f$ to facilitate comparison with  experimental spectra. Insets show the localisation of the resonant dynamics for the S and E modes with lighter shading indicating a stronger dynamic magnetisation component perpendicular to $\hext$.  The anti-dot boundary is shown as a blue circle.  (d) Schematic showing the spatial localisation of the E and S modes together with a MNP and a sketch of its stray magnetic field.}
	\label{fig2}
\end{figure}

An experimental FMR spectrum for a bare  MC (ie.~no adsorbed MNPs) obtained at 11.5 GHz  is shown in Fig.~\ref{fig2}a in which two high amplitude resonances near the extremes of the measured $\hext$ range can be identified. The frequency ($f$) dependence of the resonance fields, $\hres$, of these two FMR modes compare well with those predicted from time domain simulations for the extended (E) and side (S)  modes\cite{Neusser2008,Tacchi2010,Zivieri2013} (Fig.~\ref{fig2}b).  In Fig.~\ref{fig2}c we show Fourier transformed data from the time domain simulation obtained at $\mu_0\hext=200$ mT, showing the E and S modes  together with their spatial localisations inside the MC's unit cell. The E mode is concentrated in  bands between rows of anti-dots oriented perpendicular to $\hext$ while the S mode is  localised between neighboring anti-dots (see also the schematic in Fig.~\ref{fig2}d).   In both simulation and experiment, a number of lower amplitude modes  lie between the side and extended modes, demonstrating good agreement in terms of the overall mode structure however these modes will not be discussed in this communication. 

\begin{figure}[htbp]
\centering
	\includegraphics[width=7cm]{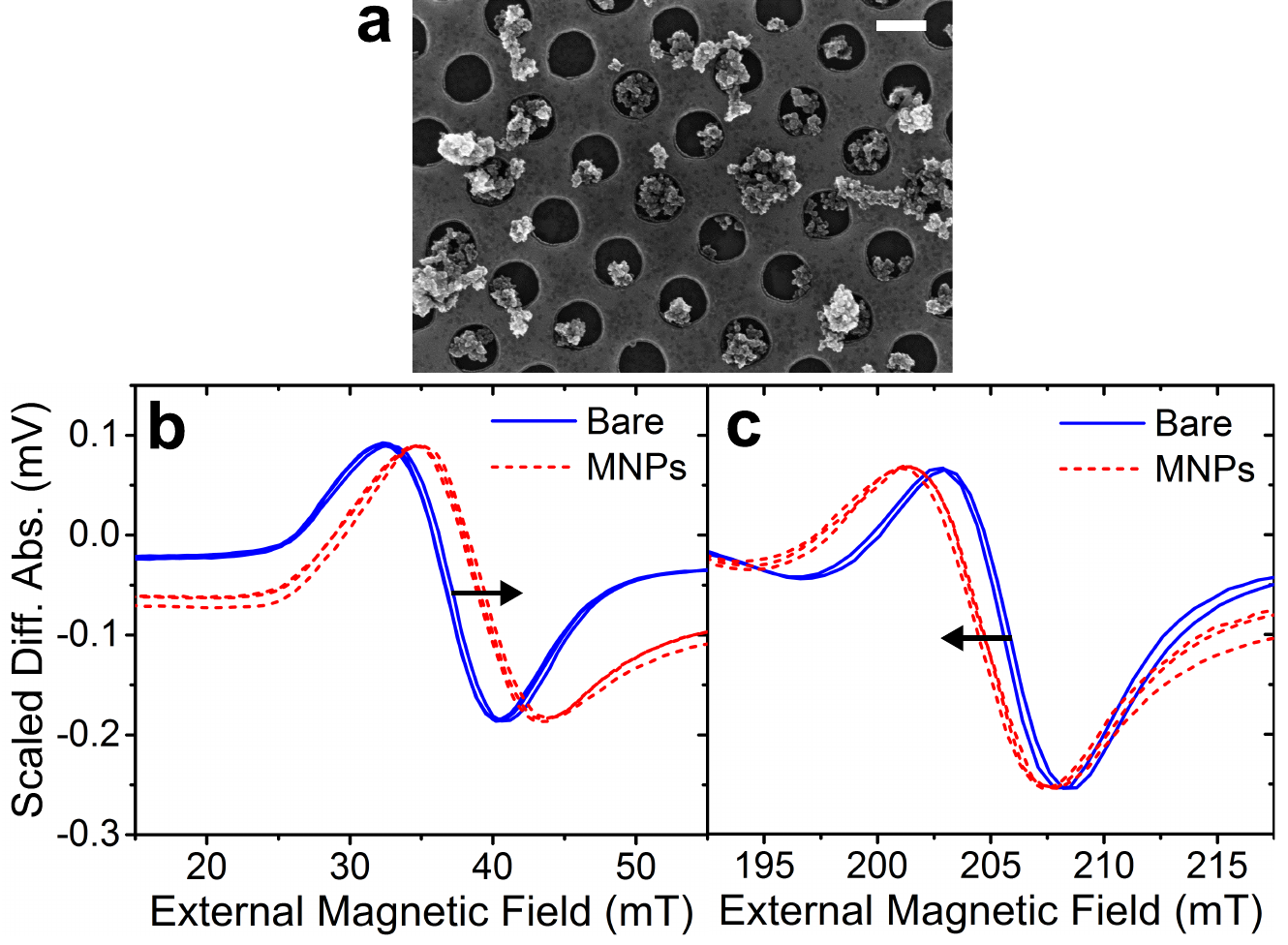}
	\caption{(a) SEM image of MNPs on the MC with a 300 nm wide scale bar. Out of 33 holes counted, 30 contain MNPs. In 7 of these, the `captured' MNP extends outside of the anti-dot. There are 5 isolated nanoparticles on the MC's upper surface.  FMR traces obtained at 11.5 GHz showing the side  mode (b) and extended mode (c) before (`bare') and after the addition of MNPs. Multiple traces have been taken following repeated removal and replacement of the MC, a requirement for MNP application, enabling an estimation of the associated experimental uncertainty in ($\sim 0.5$ mT). Traces have been vertically scaled and vertically shifted to locally normalize the differential absorption signals to that obtained with a bare MC.}
	\label{shift}
\end{figure}

MNPs were adsorbed onto the MCs at  $\hext=0$ by placing 12 $\mu$L of diluted MNP suspension on the MC's surface which then was allowed to dry in ambient conditions before re-measurement. Scanning electron microscopy of  dried MNPs on a MC   reveals irregularly shaped MNPs  with the majority lying inside the anti-dots (see Fig.~\ref{shift}a and caption).  The MNPs generate an upward shift in $\hres$ for the side mode resonance (Fig.~\ref{shift}b) and a downward shift for the extended mode resonance (Fig.~\ref{shift}c). Both shifts exceed experimental uncertainty (see caption of Fig.~\ref{shift}).

The observed shifts can be qualitatively understood by considering the field generated by an idealized magnetized MNP, located at the center of an anti-dot and magnetized along $\hext$ (Fig.~\ref{fig2}d). Treating the MNP as a dipole, the $y$-component of its stray magnetic field will, to a first approximation, reinforce $\hext$ at E mode region but oppose $\hext$  at the  S mode region. Thus, for a given experimental measurement frequency, a larger $\hext$ must be applied to attain the side mode resonance condition when MNPs are within the anti-dots. Similarly, the extended mode will be observed at a lower external field. These predictions are consistent with the experimental results (Fig.~\ref{shift}b,c) as well as analogous numerical results obtained for anti-ring structures\cite{Ding2013}.

To verify these arguments more rigorously, simulations were repeated with a 150nm wide spherical MNP within the anti-dot (see supplementary information for further details including the $\hext$-dependent MNP moment). The time domain simulation result for the MC in the presence of a MNP with $\mu_0 \hext=200$ mT is shown as a red dotted line in Fig.~\ref{fig2}c. The S mode's frequency is indeed decreased, consistent with that part of the MC experiencing a lower net field.  Likewise, an increased resonance frequency is predicted for the extended mode.  To extract numerical values for these shifts, eigenmode simulations were carried out at both $\mu_0 \hext= 37$ mT (to probe the side mode as per Fig.~\ref{shift}b) and at 206 mT (to probe the extended mode as per Fig.~\ref{shift}c). This yielded MNP-induced frequency shifts of +0.311 GHz for the E mode and -0.085 GHz for the S mode, in good agreement with time domain simulations. The local gradient of the $\fres$ vs.~$\hres$ data (Fig.~\ref{fig2}b) was then used to convert the frequency shifts into equivalent field shifts. This yielded $-6.9$ mT for the extended mode ($45$ GHz/T) and $+2.0$ mT for the side mode ($43$ GHz/T). Although the observed shift will depend on the MNP coverage, these simulations which assume one 150 nm wide MNP per anti-dot   correctly predict the order of magnitude and sign of the resonance field shift (Fig.~\ref{shift} where, albeit, not every hole is filled and there is a distribution of MNP sizes).

\begin{figure}[hbtp]
\centering
	\includegraphics[width=7cm]{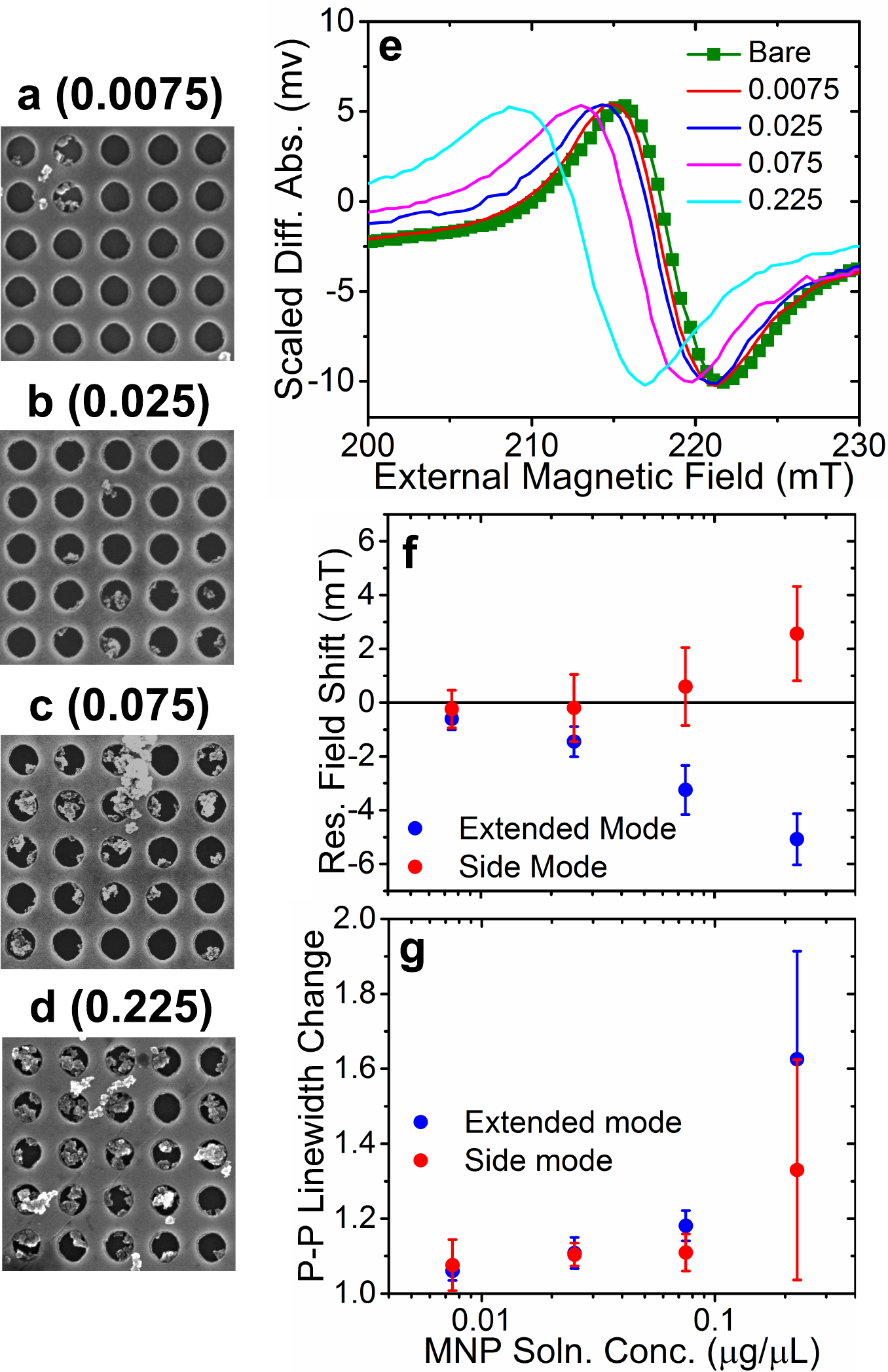}
	\caption{(a-d) 2.25 $\mu$m wide images of the MC  following consecutive applications of aqueous MNP  solutions of increasing concentrations, $c$ ($\mu$g/$\mu$L). (e) Differential absorption peak for the extended mode at 12 GHz in the bare MC and following adsorption of MNPs for increasing values of $c$. Extended and side mode peak shifts (f) and relative peak-to-peak linewidth increase (g)  as a function of MNP solution concentration.  Error bars combine both the uncertainty in the shift measurement at each frequency and the spread of peak shifts over the measured frequency range (11.5-16 GHz). }
	\label{cov}
\end{figure}

To study MNP coverage effects, we measured a second MC with an equivalent anti-dot lattice geometry. We carried out consecutive applications of diluted solutions of MNPs with increasing concentration, $c$, obtaining  FMR traces and imaging the MC via SEM before and after the application of each solution. Representative SEM images are shown in Figs.~\ref{cov}a-d for each concentration. Increases in $c$ visibly increase the MNP coverage, resulting in an increased peak shift for both the extended and side modes. For the lowest coverage, a shift of the extended mode  is measurable and just higher than the experimental uncertainty (Fig.~\ref{cov}e). At the highest  coverage however the resonance peak shift is much larger and approaches the peak-to-peak resonance linewidth. There was some variation in the measured shifts for different $f$ however  no clear, widely applicable trends could be determined (except at low $f$ for the side mode, and thus low $\hext$, where a smaller shift  presumably resulted from a lower $\hext$-induced MNP moment). In Figure~\ref{cov}f, we have averaged the S and E modes' peak shifts over the measured frequency range ($11.5-16$ GHz) and plotted the averaged shifts versus $c$. Reliably measurable shifts of the E mode are observed over the entire frequency range. However, a higher concentration must be used to register a consistent shift for the side mode. Notably, the simulated shifts of $+2$ mT for the S mode  and $-6.9$ mT for the E mode  compare well with the shifts observed at $c=0.225$ $\mu$g/$\mu$L where we are close to having every anti-dot filled with a MNP (Fig.~\ref{cov}d)) and thus closest to the simulation condition. The MNPs also generate a coverage-dependent linewidth increase with a $\sim$1.5 times increase on average in the peak-to-peak resonance linewidths at the highest $c$  (Fig.~\ref{cov}g). However, the linewidth broadening does not dominate the observed shifts in that the minimum of the differential absorption line consistently moves in the direction of the peak shift (Fig.~\ref{cov}e). This is qualitatively consistent with that expected for a collection of differently sized MNPs all acting in unison with the shift magnitude depending upon the size of the MNP.

\section*{Conclusions}

We have used an anti-dot based magnonic crystal (MC) to experimentally demonstrate the ability to detect magnetic nanoparticles (MNPs) via their influence on spatially confined, high frequency, ferromagnetic resonance modes. MNPs are preferentially captured within the holes which, depending upon a mode's spatial localisation, leads to an increased or decreased resonant frequency. Resonance shifts are reproduced well by micromagnetic simulations and observable even for quite low anti-dot fillings ($\sim 15\%$).  A non-dominant linewidth broadening is observed at high MNP coverages. Our results are directly applicable to confined modes  in \textit{isolated} spintronic\cite{Braganca2010} or magnonic\cite{Urazhdin2014} nanostructures. This is encouraging for the development of frequency-based spintronic devices such as spin torque oscillators for nano-scale magnetic biosensing\cite{Braganca2010} in applications such as flow cytometry\cite{Loureiro2011a,Helou2013}.  Notably our observed frequency shifts are  significantly larger than measured spin torque oscillator linewidths\cite{Pribiag2007,Zeng2012a,LeBrun2014} however  electrical detection will rely on close proximity of the MNP to the sensing layer and high GHz/T field sensitivities.  

\section*{Acknowledgements}

Research supported by the Australian Research Council's Discovery Early Career Researcher Award (DE120100155) and Discovery Projects scheme (DP110103980), the University of Western Australia's (UWA)  RCA,  ECRFS, SIRF, UPAIS, Vacation Scholarship, Teaching Relief and post-doctoral fellowship schemes, an EPSRC DTC grant (EP/G03690X/1) and  iVEC through the use of advanced computing resources located at iVEC@UWA. A.O.A. was supported by the National Research Foundation, Prime Minister's Office, Singapore under its Competitive Research Programme (CRP Award No. NRF-CRP 10-2012-03). The authors thank C.~Lueng, D.~Turner, A.~Suvorova,  A.~Vansteenkiste, A.~Dodd,  M.~House, T.G.~St.~Pierre, N.~Kostylev, C.~Bording and J.~Izaac  for useful discussions, advice and/or assistance. The authors  acknowledge access to the UWA's Biomagnetics Wet Laboratory and Magnetic Characterisation Facility as well as the facilities, and the scientific and technical assistance of the Australian Microscopy \& Microanalysis Research Facility at the Centre for Microscopy, Characterisation \& Analysis, The University of Western Australia, a facility funded by the University, State and Commonwealth Governments. 

\section*{Supplementary information}

\subsection*{Magnonic crystals and magnetic nanoparticles } 

\begin{figure}[htbp]
\centering
	\includegraphics[width=7cm]{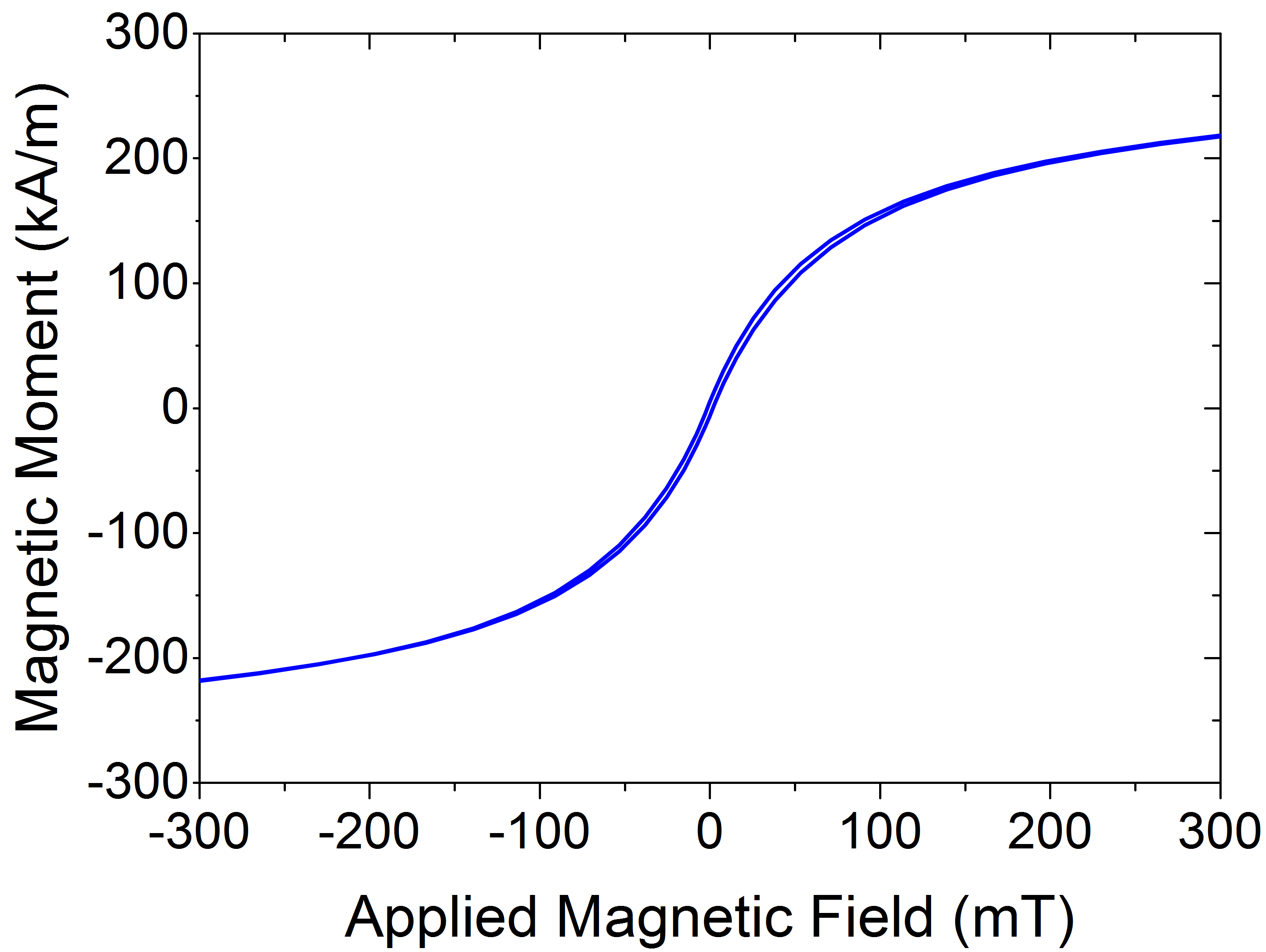}
	\caption{[Supplementary figure] Magnetic moment per unit volume for the MNPs at 300K. }
	\label{figsquid}
\end{figure}

The MCs were composed of 30 thick nm Ni$_{80}$Fe$_{20}$ layers covered by an Au capping layer (8 nm thick for the data in Manuscript Figs.~2 and 3  and 10 nm thick for the data  in Manuscript Fig.~4). They were fabricated on  Si substrates using deep ultraviolet lithography with deposition via electron beam evaporation followed by lift-off\cite{Singh2004,Adeyeye2008}. The antidot lattice geometries  were  measured using scanning electron microscopy (SEM)   with antidot diameters and array pitch rounded to the nearest 10 nm. MNPs were nanomag-D iron-oxide nanoparticles (micromod Partikeltechnologie GmbH) with a colloidally stabilized dextran surface, a quoted MNP width of 130 nm and a quoted solids content of 25 mg/mL. The latter was used to calculate the concentrations of the diluted MNP solutions (diluted using e-pure water). Magnetometry on freeze dried MNPs  was carried out at 300K using a MPMS3 SQuID magnetometer (Quantum Design Inc.) in VSM mode. The magnetic moment per unit volume (Fig.~\ref{figsquid}) was calculated assuming an iron-oxide density of 5.24 g/cm$^3$. SEM was carried out with a Zeiss 1555 VP-FESEM and a FEI Verios 460 SEM.

\subsection*{Ferromagnetic resonance spectroscopy } 

FMR modes were probed at room temperature using broadband, stripline-based, magnetic field modulated FMR spectroscopy (Manuscript Fig.~1b). During measurement, the MC's (01) axis was closely aligned with stripline which itself was aligned with $\hext$, the latter modulated at 220 Hz. The technique  uses a interferometric receiver\cite{Ivanov2014} and lock-in amplifier (SR830, Stanford Research Systems) to measure the external magnetic field ($\hext$-)derivative of finite width ferromagnetic resonances in the sample at a set frequency, $f$, and stepped $\hext$.  To obtain the FMR traces, $\hext$ was increased in steps over a range typically spanning $\sim 0-250$ mT with $\hext$ measured at each step using a FH54 Teslameter  (Magnet-Physik Dr.~Steingroever GmbH). The resultant FMR spectra were measured both before and after the addition of cluster-shaped  MNPs to the MC's surface. The MCs are characterised by a non-zero remanent magnetisation with the stepped $\hext$ sweeps for the measurements carried out for a single field polarity. Thus, the bias field and the spatially averaged $y$-component of the MC's magnetic  moment remained aligned during the FMR  experiments.    A microscope coverslip between the MC and the stripline was used in all measurements to avoid MNPs, when present, rubbing off onto the stripline. A PVC block placed on the stripline board without contact to the stripline itself was used to ensure consistent placement of the sample with respect to the stripline. This enabled excellent reproducibility even with repeated removals and replacements of the MC. Reproducibility was confirmed for each measurement with the associated uncertainty in the peak position  typically being on the order of 0.5 mT. This can be seen in Manuscript Figs.~3b,c where we show traces obtained after repeated removals and replacements of the sample.   Upon adding MNPs to the MC, we consistently observed an overall decrease in the signal amplitude which was stronger than the MNP-induced linewidth broadening effects. This signal amplitude reduction increased with MNP concentration, suggesting a broadband  absorption of the microwave power by the MNPs. To consistently compare data obtained with and without MNPs, we vertically scaled the FMR data so that all traces had the same peak-to-peak amplitude. It was also sometimes necessary to introduce a small vertical offset, typically on the order of a few tens of $\mu$V at most.

\subsection*{Micromagnetic simulations} 

Two micromagnetic simulation methods were used in this work. Both methods simulated a single unit cell of the MC with periodic boundary conditions in the $x$ and $y$ directions, an $11\times 11$ `tiled' macro-geometry\cite{Fangohr2009} for determining the demagnetizing field and the following magnetic parameters: damping $\alpha=0.008$, nil intrinsic anisotropy, saturation magnetization $M_S=8\times 10^5$ A/m, gyromagnetic ratio $2\pi\gamma=1.85\times 10^{11}$ rad/T.s and exchange stiffness $A_{ex}=13$ pJ/m. These values of $M_S$ and $\gamma$ were consistent with results from FMR measurements on continuous layers averaged over two reference samples. The value of $\gamma$ is also close to that determined by Shaw \textit{et al}\cite{Shaw2013}. MNPs were modeled explicitly assuming a ferromagnetic sphere sitting within the anti-dot with its lower surface aligned with the lower surface of the MC. For the MNP, we used $\alpha=0.05$, $A_{ex}$ and $\gamma$ equal to that in the MC. $M_S$ was read off from Fig.~\ref{figsquid} at the value of $\hext$ used in the simulation. IPython\cite{Perez2007}, Sumatra\cite{Davison2012} and matplotlib\cite{Hunter2007} were used for analysis, management and visualisation of the simulation data.

\medskip

\noindent \textbf{Time domain (ringdown) simulations }  Time domain or `ringdown' simulations (e.g.~Grimsdith \textit{et al.}~\cite{Grimsditch2004}) were carried out using MuMax3\cite{Vansteenkiste2014} version 3.5.3 with cuboid discretisation cells ($\approx$ $3.52 \times 3.52 \times 3.75$ nm$^3$) wherein the system's equilibrium magnetic configuration, $\mathbf{m_0}(\mathbf{r},\hext)$, at a given $\hext$, applied in the $y$-direction, is perturbed with a  field pulse in the $x$-direction (0.5 mT sinc pulse\cite{Venkat2013} with a 300 ps offset and 30 GHz cut-off frequency). Fourier analysis was then applied to the time dependent, spatially averaged $x$-component of the magnetization to extract the characteristic (resonant) frequencies associated with the resultant excited dynamics. Mode visualisations (insets of Manuscript Fig.~2c) were obtained by calculating the spatially resolved Fourier amplitudes for $m_x$ at each  resonant frequency.  

\medskip

\noindent \textbf{Eigenmode simulations }  The second method was an  eigenmode calculation which uses the finite element micromagnetic simulator, FinMag (successor to Nmag\cite{Fischbacher2007}), to directly determine the eigenfrequencies and eigenvectors associated with a given $\mathbf{m_0}(\mathbf{r},\hext)$.  For relaxation, the system was meshed using a characteristic internode length of $\lmesh=3.5$ nm. The eigenmodes were determined  using an algorithm similar to that detailed in d'Aquino \textit{et al.}\cite{dAquino2009}~on a coarsened mesh with $\lmesh=7$ nm (coarsening was needed due to the algorithm's high memory requirements). The nature of the mode (e.g.~side or extended) was determined via visual inspection of the eigenvectors. 
 
\providecommand*{\mcitethebibliography}{\thebibliography}
\csname @ifundefined\endcsname{endmcitethebibliography}
{\let\endmcitethebibliography\endthebibliography}{}

\end{document}